\documentclass[]{aa} 
\usepackage{graphicx}
\usepackage{amsmath, mathtools}
\usepackage{txfonts}
\usepackage{color,ulem,lscape, ascmac, bm}
\newcommand{\sub }[1]{_{\mathrm{#1}}}
\newcommand{\Mstar}{M_{\star}}
\newcommand{\Lstar}{L_{\star}}

\newcommand{\Teff}{T\sub{eff}}
\newcommand{\Mdot}{\dot{M}\sub{acc}}

\newcommand{\Msun}{  M_\odot }
\newcommand{\Rsun}{  R_\odot }
\newcommand{\Lsun}{  L_\odot}

 \newcommand{ \Dmix }{ D_\mathrm{mix} }
 \newcommand{ \tmix }{ t_\mathrm{mix} }
 \newcommand{ \DT }{ D_T }
 
 \newcommand{ \Ys }{ Y\sub{surf} }
 \newcommand{ \Zs }{ Z\sub{surf} }
 \newcommand{ \ZXs }{ (Z/X)\sub{surf} }
 \newcommand{ \Zc }{ Z\sub{center} }
 \newcommand{ \Xc }{ X\sub{center} }
 \newcommand{ \Yc }{ Y\sub{center} }
 \newcommand{ \Tc }{ T\sub{center} }
 \newcommand{ \muc }{ \mu\sub{center} }

 \newcommand{ \Zacc }{ Z\sub{accretion} } 
 \newcommand{ \Xaccini }{ X\sub{proto} } 
 \newcommand{ \Yaccini }{ Y\sub{proto} } 
  
 \newcommand{ \Zaccini }{ Z\sub{proto} }
 \newcommand{ \Zaccmax }{ Z\sub{acc,max} }
 \newcommand{ \cs }{ c_\mathrm{s} }
 \newcommand{ \csobs }{ c_\mathrm{s,obs} }
 \newcommand{ \delcs }{ \delta \cs }
 \newcommand{ \fopa }{ \delta_\kappa }
 \newcommand{ \RCZ }{ R\sub{CZ} }

 \newcommand{ \Yproto }{Y\sub{proto}}
 \newcommand{ \Zproto }{Z\sub{proto}}
 \newcommand{ \chitwo }{ \chi^2 }
 
 \newcommand{ \fov }{ f\sub{overshoot} }
 \newcommand{ \amlt }{ \alpha\sub{MLT} }

\newcommand{ \ALisun}{ A(^7\mathrm{Li})_\odot }
\newcommand{ \ABesun}{ A(^9\mathrm{Be})_\odot }

\newcommand{ \ALi}{ A(^7\mathrm{Li}) }
\newcommand{ \ABe}{ A(^9\mathrm{Be}) }
\newcommand{ \phiB}{ \varPhi(\element[][8]{B}) }
\newcommand{ \phiBe}{ \varPhi(\element[][7]{Be}) }

\newcommand{ \phipp}{ \varPhi( {pp} ) }
\newcommand{ \phipep}{ \varPhi( {pep} ) }
\newcommand{ \phiCNO}{ \varPhi( \mathrm{CNO} ) }
\defcitealias{Kunitomo+Guillot21}{KG21}
\defcitealias{Kunitomo+22}{KGB22}
\defcitealias{GS98}{GS98}
\defcitealias{Asplund+09}{AGSS09}
\defcitealias{Asplund+21}{AAG21}

\usepackage[colorlinks = true,
        linkcolor = blue,
        urlcolor  = blue,
        citecolor = blue,
        anchorcolor = blue
]{hyperref}
\makeatletter
\renewcommand*\aa@pageof{, page \thepage{} of \pageref*{LastPage}}
\makeatother

\bibpunct{(}{)}{;}{a}{}{,} 

\begin{document} 

    \title{Solar models with protosolar accretion and turbulent mixing}

   \author{ \href{https://orcid.org/0000-0002-1932-3358}{Masanobu Kunitomo}\inst{1,2}
          \and
            \href{https://orcid.org/0000-0001-6357-1992}{Ga\"{e}l Buldgen}\inst{3}
          \and
          \href{https://orcid.org/0000-0002-7188-8428}{Tristan Guillot}\inst{2}
          }

   \institute{Department of Physics, Kurume University, 67 Asahimachi, Kurume, Fukuoka 830-0011, Japan\label{inst1}\\
		\email{kunitomo.masanobu@gmail.com}
        \and
	Universit\'e C\^ote d'Azur, Observatoire de la C\^ote d'Azur, CNRS, Laboratoire Lagrange, Bd de l'Observatoire, CS 34229, 06304 Nice cedex 4, France\label{inst2}
   		\and
        STAR Institute, Universit\'{e} de Li\`{e}ge, Li\`{e}ge, Belgium\label{inst3}
}
   \date{Received 10 April 2025 / Accepted 21 August 2025}
   \authorrunning{M. Kunitomo, G. Buldgen \& T. Guillot}

 
  \abstract
{
Over the last two decades, standard solar models (SSMs) have never reproduced all the observational data, resulting in active discussions on the so-called ``solar modeling problem.'' A recent study suggested that the accretion from the protosolar disk onto the proto-Sun can leave a large compositional gradient in the solar interior, in agreement with the low-metallicity ($Z$) solar surface and the high-$Z$ solar core suggested by spectroscopic and neutrino observations, respectively. In addition, recent analyses have reported low lithium but high beryllium abundances on the solar surface; however, SSMs predict Li abundances that are $\sim$30$\sigma$ away from the observed value.
}{
In this study, we aim to develop solar models and compare them with the Li and Be abundance constraints.
}{
We examine the effect of accretion and turbulent mixing below the base of the surface convective zone. We compute $\sim$200 solar evolutionary models for each case to optimize input parameters using target quantities, similar to the SSM framework.
}{
We confirm that turbulent mixing helps reproduce the surface Li and Be abundances within $\sim$0.6$\sigma$ by enhancing burning. It suppresses gravitational settling, leading to a better matching of the He surface abundance ($\lesssim$0.3$\sigma$) and a smaller compositional gradient. We derive a new protosolar helium abundance $\Yproto=0.2651 \pm 0.0035$.
Turbulent mixing decreases the central metallicity ($\Zc$) by $\approx$4.4\%, even though accretion increases $\Zc$ by $\approx$4.4\%, as suggested by our previous study.
Unfortunately, the reduction in $\Zc$ implies that our models do not reproduce constraints on observed neutrino fluxes by $6.2\sigma$ for \element[][8]{B} and $2.7\sigma$ for CNO.
}{
Including turbulent mixing in solar models appears indispensable to reproduce the observed atmospheric abundances of Li and Be. However, the resulting tensions in terms of neutrino fluxes, even in the models with the protosolar accretion, show that the solar modeling problem remains, at least partly. We suggest that improved electron screening, as well as other microscopic properties, may help alleviate this problem. An independent confirmation of the neutrino fluxes measured by the Borexino experiment would also be extremely valuable.
}


   \maketitle
%

\section{Introduction}
\label{sec:intro}

The Sun is a benchmark star for stellar structure and evolution theory. Much effort has been put into spectroscopic, helioseismic, and neutrino observations for a long time. These constraints have been used to test theoretical models. Over the last two decades, standard solar models \citep[SSMs hereafter; ][]{Christensen-Dalsgaard+96}, which are solar models constructed with standard input physics, do not reproduce the surface metallicity ($\Zs$), sound speed profile, neutrino fluxes simultaneously \citep[][]{Vinyoles+17, Buldgen+19, Christensen-Dalsgaard21}. This so-called solar modeling problem (or solar abundance problem) has been actively discussed in the community, with a clear solution yet to be found.

Concerning the solar modeling problem, two processes have drawn attention. The fit to the helioseismic constraints of the sound speed ($\cs$) profile at the base of the surface convective zone (BCZ hereafter) can be significantly improved by an increase in opacity by $\sim$10\% \citep[e.g.,][]{Christensen-Dalsgaard+09, Villante10, Bailey+15, Buldgen+19, Kunitomo+Guillot21}.
In addition, while SSMs do not include the protosolar phase, the proto-Sun grew by accretion from the protosolar disk, where the planets were formed. Planet formation theory predicts that dust grains drift rapidly in the disk and thus the dust-to-gas ratio (i.e., metallicity) of the accretion flow onto the proto-Sun must have been variable \citep[see][hereafter \citetalias{Kunitomo+Guillot21}]{Kunitomo+Guillot21}. The variable metallicity of accretion $\Zacc$ (in particular, the low metallicity of the final phases of accretion) implies that the metallicity of the proto-Sun and thus the central metallicity ($\Zc$) of the present-day Sun are larger than usually assumed (\citetalias{Kunitomo+Guillot21}). The higher $\Zc$ affects the thermal structure of the solar core, and thus its neutrino fluxes. \citet[][hereafter \citetalias{Kunitomo+22}]{Kunitomo+22} demonstrated for the first time that the models with an ad hoc opacity increase and variable $\Zacc$ reproduce all three constraints ($\Zs$, $\cs$, and neutrinos) simultaneously.

Recently, the abundance of light elements (i.e., lithium and beryllium) has been the subject of several studies \citep[][]{Amard+16, Eggenberger+22, Buldgen+25a, Yang+25}. These elements are burned at relatively low temperatures, $\sim$2.5\,million K (MK hereafter) for Li and $\sim$3.5\,MK for Be, and thus their abundances are tied to the internal mixing process.
Compared to the meteoritic constraints for the Solar-System primordial abundances \citep[$\ALi=3.27\pm0.03$\,dex and $\ABe=1.31\pm0.04$\,dex, respectively; ][]{Lodders21}\footnote{The abundance of an element X is $A(\mathrm{X})\equiv\log\left( 
N_\mathrm{X}/N_\mathrm{H} \right) +12 $, where $N_\mathrm{X}$ is the number density. Throughout this paper, $\log\equiv\log_{10}$.}, recent analyses indicate that, at the surface of the present-day Sun, Li is depleted by $>$2\,dex \citep[$\ALisun=0.96$\,dex; ][see Table\,\ref{tab:targets}]{Wang+21} whereas Be is only slightly depleted by $\sim$0.1\,dex \citep[$\ABesun=1.21$\,dex; ][]{Amarsi+24}. Thus, in the solar interior, a mixing process should operate beyond the Li-burning region ($\lesssim$0.65\,$\Rsun$) but should not reach the Be-burning region ($\sim$0.55\,$\Rsun$).

\citet{Eggenberger+22} have shown that the Li abundance and rotation profile can be reproduced by rotational mixing; namely, rotation-induced hydrodynamic and magnetohydrodynamic instabilities such as circulation, shear mixing, and magnetic Tayler instabilities \citep[see also][]{Buldgen+24b}. 
This is not the case for SSMs, which predict a value $30\sigma$ away from the observed $\ALisun$ \citep[see Fig.\,6 of][]{Eggenberger+22}.
However, the rotational mixing in \citet{Eggenberger+22} is too deep to reproduce $\ABesun$. Another implementation of rotational mixing has also been examined by \citet{Yang+25}, but the latest values of $\ALisun$ and $\ABesun$ have not been reproduced simultaneously \citep[][]{Wang+21, Amarsi+24}.
Instead, a more localized shallow turbulent mixing is preferred \citep[][]{Buldgen+25a}.

Therefore, the question of the present article is: If we consider protosolar accretion, an increase of the interior opacities, and turbulent mixing, can we reproduce $\Zs$, $\cs$, neutrino fluxes, and the Li and Be abundances simultaneously?
This paper is organized as follows: we describe the computation method in Sect.\,\ref{sec:model}, show the results (evolution of surface Li and Be abundances, metallicity profile at the solar age, and neutrino fluxes) in Sect.\,\ref{sec:results}, discuss the origin of turbulent mixing and various effects on neutrino fluxes in Sect.\,\ref{sec:discussions}, and results are summarized in Sect.\,\ref{sec:conclusion}.

\begin{table}
\caption{Observational constraints of the present-day Sun.}
\label{tab:targets}
\centering
\begin{tabular}{lllll}
\hline\hline\noalign{\smallskip}
Parameter & Value & Uncertainty & Unit & Ref. \\ 
\hline
  $\log\Lstar$ & 0 & 0.01\,dex & $\Lsun$  & 1, 2 \\
  $\Teff$ & 5777 & 10 & K & 1, 2 \\
  $\ZXs$   & {\bf 0.0187} & {\bf 0.0009} & & {\bf 3} \\
    \noalign{\smallskip} \hline \noalign{\smallskip}
  $\Ys$ & 0.2485 & 0.0035 & & 4 \\
  $\RCZ$  & 0.713 & 0.01 & $\Rsun$ & 1, 2\\
  rms($\delcs$) & 0 & $10^{-3}$ & & 1, 7\\
    \noalign{\smallskip} \hline \noalign{\smallskip}
  $\ALisun$ &  {\bf 0.96} & {\bf 0.05} & dex & {\bf 5} \\
  $\ABesun$ &  {\bf 1.21} & {\bf 0.05} & dex & {\bf 6} \\
    \noalign{\smallskip} \hline \noalign{\smallskip}
  $\phiB$ &  5.16 & $^{+0.13}_{-0.09}$ & $10^{6}\,\mathrm{s^{-1}cm^{-2}}$ & 8 \\ \noalign{\smallskip}
  $\phiBe$ &  4.80 & $^{+0.24}_{-0.22}$ & $10^{9}\,\mathrm{s^{-1}cm^{-2}}$ & 8 \\ \noalign{\smallskip}
  $\phipp$ &  5.971 & $^{+0.037}_{-0.033}$ & $10^{10}\,\mathrm{s^{-1}cm^{-2}}$ & 8 \\ \noalign{\smallskip}
  $\phipep$ &  1.448 & $^{+0.013}_{-0.013}$ & $10^{8}\,\mathrm{s^{-1}cm^{-2}}$ & 8 \\ \noalign{\smallskip}
  $\phiCNO$ &  {\bf 6.7} & $\bm{^{+1.2}_{-0.8}}$ & $10^{8}\,\mathrm{s^{-1}cm^{-2}}$ & {\bf 9} \\ \noalign{\smallskip}
\hline
\end{tabular}
\tablefoot{
In the SSM framework, only the first three are used as target values.
The bold texts highlight updates from \citetalias{Kunitomo+Guillot21} and \citetalias{Kunitomo+22}. See Appendix\,\ref{sec:abundance} for the discussion on the $\Ys$ constraint.
The solar age is 4.567\,Gyr \citep{Amelin+02}.
\tablebib{
(1) \citetalias{Kunitomo+Guillot21},
(2) \citet{Bahcall+05},
(3) \citetalias{Asplund+21}, 
(4) \citet{Basu+Antia04},
(5) \citet{Wang+21},
(6) \citet{Amarsi+24},
(7) \citet{Basu+09},
(8) \citet{Orebi-Gann+21},
(9) \citet{Basilico+23}.
}
}
\end{table}

\section{Model} \label{sec:model}

\begin{figure*}[ht]
  \begin{center}
    \includegraphics[angle=0,width=0.8\hsize]{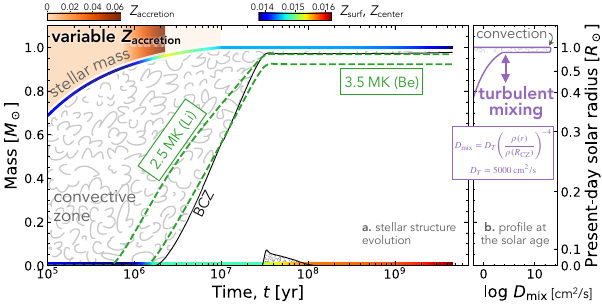}
  \end{center}
  \caption{
  Schematic illustration showing the structure evolution (panel a; so-called Kippenhahn diagram) and the $\Dmix$ profile at the solar age (panel b) of our fiducial model ``K2-MZvar-TM'' (see Table\,\ref{tab:chi2}). The convective zones are depicted as a cloudy region. The color of the shade in panel a shows $\Zacc$. The two lines at the stellar surface and center also illustrate metallicity by color. The two green dashed lines indicate the locations at temperatures $T=2.5\times10^6$ and $3.5\times10^6$\,K, indicative of Li and Be burning, respectively.}
  \label{fig:schematic}
\end{figure*}

\begin{table*}[!t]
	\begin{center}
	\caption{Settings of optimization simulations.}
	\label{tab:chi2}
           \begin{tabular}{lccccccl}
            \hline
            \hline
            \noalign{\smallskip}
            Model name & Abundance & $\Zacc$ & $n$ & $\DT\,[\mathrm{cm^{2}s^{-1}}]$ & $\fov$ & $A_2$ [\%] & Comment\\
            \noalign{\smallskip}
            \hline
            \noalign{\smallskip}
            K2 & \citetalias{Asplund+21} &steady & --- & --- & --- & 10 &  \\
            K2-MZvar & \citetalias{Asplund+21} &variable & --- & --- & --- & 10 & \\
            K2-TM & \citetalias{Asplund+21} &steady & 4 & 5000 & --- & 10 & \\
            \textbf{K2-MZvar-TM} & \citetalias{Asplund+21} & variable & 4 & 5000 & --- & 10 & \textbf{Fiducial} \\
            \noalign{\smallskip}
            \hline
            \noalign{\smallskip}
            K2-A2 & \citetalias{Asplund+21} & steady & --- & --- & --- & [0, 20] & Appendix\,\ref{sec:abundance} \\
            K2-fov & \citetalias{Asplund+21} & steady & --- & --- & 0.01/0.023 & 10 & Appendix\,\ref{sec:ovs} \\
            \noalign{\smallskip}
            \hline
            \noalign{\smallskip}
            K2-A09 & \citetalias{Asplund+09} & steady & --- & --- & 0.0103 & 12 & $N=6$; from \citetalias{Kunitomo+22} \\
            K2-MZvar-A09 & \citetalias{Asplund+09} & variable & --- & --- & 0.0042 & 12 & $N=6$; from \citetalias{Kunitomo+22} \\
            \hline
            \end{tabular} 
         \end{center}
         \tablefoot{
         All the new models in this study use three input parameters and three target values ($N=M=3$).
         The parameters $n$ and $\DT$ control turbulent mixing (see Eq.\,\ref{eq:Dmix}).
         The ``K2-A09'' and ``K2-MZvar-A09'' models correspond to ``K2-A2-12'' and ``K2-MZvar-A2-12'' in \citetalias{Kunitomo+Guillot21} and \citetalias{Kunitomo+22}: the names are changed to emphasize the abundances.}
\end{table*}

We simulate solar models with the Modules for Experiments in Stellar Astrophysics (MESA) stellar evolution code (version 12115), including the effects of protosolar accretion and turbulent mixing. For details, we refer the reader to the Paxton et al. papers and our previous papers \citep[][\citetalias{Kunitomo+Guillot21}, and \citetalias{Kunitomo+22}]{Kunitomo+17, Kunitomo+18}. Below, we summarize the computational method and the updates in this study. We mainly focus on protosolar accretion, turbulent mixing, opacity increase, abundance scale, and optimization.

Figure\,\ref{fig:schematic} illustrates two crucial physical processes of this paper, namely protosolar accretion with variable composition and turbulent mixing in the radiative zone. The proto-Sun grew by accretion from the protosolar disk, where planets were formed. Planets were formed by the coagulation of dust grains in the disk. Once the dust grains became cm-sized, which are called pebbles, they rapidly migrated onto the proto-Sun due to the frictional force with the disk gas. This led to an increase of the metallicity of the accreted gas, $\Zacc$. We call this the ``pebble wave phenomenon.'' In contrast, in the late phase, the disk gas became depleted in heavy elements/metals due to the drift or the filtration by proto-giant planets \citep{Guillot+14}. Thus, a variable composition of the accreted gas is a natural consequence of recent planet formation theory (see \citetalias{Kunitomo+Guillot21} and references therein). We also compute models with steady $\Zacc$ (i.e., constant primordial metallicity, $\Zproto$, over time), which correspond to the cases with no planet formation processes.

We adopt a $\Zacc$ model following \citetalias{Kunitomo+Guillot21} (see their Fig.\,4): $\Zacc=\Zproto$ in the earliest phase (until stellar mass, $\Mstar$, reaches $M_1$), $\Zacc$ increases up to $\Zaccmax$ during the pebble-wave phase ($M_1<\Mstar<M_2$), and then $\Zacc=0$ in the late phase. For simplicity, in the models with variable $\Zacc$, we adopt $M_1$, $M_2$, and $\Zaccmax$ values from the best model K2-MZvar-A2-12 in \citetalias{Kunitomo+Guillot21} ($M_1=0.90\,\Msun$, $M_2=0.96\,\Msun$, and $\Zaccmax=0.065$; see their Table A.1). Simulations start with a $0.1\,\Msun$ seed, mass accretion rate, $\Mdot$, decreases with time following \citet{Hartmann+98}, and accretion stops at 10\,Myr (see Fig.\,3 of \citetalias{Kunitomo+Guillot21}).

For turbulent mixing, we follow \citet{Proffitt+Michaud91} and \citet{Buldgen+25a}; thus, the diffusion coefficient in the radiative core in the main sequence (MS hereafter) is given by
\begin{align} \label{eq:Dmix}
    \Dmix=D_T\left(\frac {\rho(r)} {\rho \left(\RCZ \right)} \right)^{-n}
\end{align}
where $\rho$ is the density, $r$ is the radius, $\RCZ$ is the radius at the BCZ, $\DT$ is the diffusion coefficient at the BCZ, and $n$ is a power-law index. We turn on mixing at 30\,Myr \footnote{We consider mixing only in the MS following previous studies \citep[][]{Proffitt+Michaud91, Buldgen+25a} but we confirmed that the choice of this starting time does not affect the conclusion of this study.}.
\citet[][see their Table\,3]{Buldgen+25a} investigated a wide range of $n$ and $\DT$ sets and found that $\ALisun$ and $\ABesun$ can be well reproduced if $n\sim3$--6 and $\DT\sim3000$--$10000\,\mathrm{cm^2/s}$. In this study, we adopt $n=4$ and $D_T=5000\,\mathrm{cm^2/s}$ for the models including mixing.

As for the other internal mixing processes, convection is treated by the mixing-length theory \citep[][]{Cox+Giuli68}. Overshooting is not considered except for models ``K2-fov'', as in \citetalias{Kunitomo+Guillot21} and \citetalias{Kunitomo+22} (see Appendix\,\ref{sec:ovs}).
Gravitational setting is included \citep[][]{Thoul+94}\footnote{
Elements are lumped into four groups and the diffusion velocity is calculated only for four representative elements, namely \element[][1]{H}, \element[][4]{He}, \element[][16]{O}, and \element[][56]{Fe} \citep[see Sect.\,5.4 of][]{Paxton+11}.
}. Radiative levitation is not considered.

The opacity in the solar BCZ condition remains uncertain and is actively discussed. As in \citetalias{Kunitomo+Guillot21}, we consider opacity increase at around the BCZ as
\begin{align}\label{eq:kap}
    \kappa' &= \kappa (1 + \fopa)\,,
\end{align}
where\footnote{The opacity increase function in this study is a single Gaussian function and we do not consider other opacity increases (i.e., $A_1=0$ and $A_3=0$ in \citetalias{Kunitomo+Guillot21}).}
\begin{align}\label{eq:delkap}
    \fopa = A_2\,\exp \left[ -\frac{ \left(  \log (T/{\rm K}) -b_2\right)^2}{2c_2^2} \right]\,.
\end{align}
We adopt $b_2=6.45$ and $c_2=0.18$ \citep[][see also Fig.\,\ref{fig:A2}]{LePennec+15} and $T$ is the temperature. We determined the free parameter $A_2$ to be 10\% from a parameter study with the ``K2-A2'' models and adopted this value for all other models.
We note that this value is similar to 12\% used in \citetalias{Kunitomo+Guillot21} and \citetalias{Kunitomo+22} (see Appendix\,\ref{sec:abundance}), and close to the suggestions by experiments \citep[4--10\%;][]{Bailey+15} and helioseismic analyses \citep[$\sim$10\% at $\sim$2\,MK; ][]{Buldgen+25b}.
We refer the reader to \citetalias{Kunitomo+Guillot21} and references therein for more details.

In this study, we adopt the solar abundance scale in \citet[][hereafter \citetalias{Asplund+21}]{Asplund+21}, instead of \citet[][hereafter \citetalias{Asplund+09}]{Asplund+09} used by \citetalias{Kunitomo+Guillot21} and \citetalias{Kunitomo+22}. Therefore, we have used the different target $\ZXs$ value, corresponding opacity tables, $A_2$ value, and initial seed, from \citetalias{Kunitomo+Guillot21} and \citetalias{Kunitomo+22} (see Table\,\ref{tab:targets} and Appendix\,\ref{sec:abundance} for more details).

Table\,\ref{tab:chi2} summarizes the settings of the simulation models under a variety of settings (e.g., with and without turbulent mixing, variable or steady $\Zacc$). We optimize input parameters using the downhill simplex method \citep[][]{Nelder+Mead65}. In this study, we use $\ZXs$, $\Lstar$, and $\Teff$ (surface abundance ratio of metals to hydrogen, luminosity, and effective temperature, respectively; see Table\,\ref{tab:targets}) as target values in order to iteratively calibrate $\amlt$, $\Yproto$, and $\Zproto$ (mixing-length parameter, primordial helium abundance, and primordial metallicity, respectively). Thus, $N=M=3$, where $N$ and $M$ are the numbers of target values and input parameters, respectively. In optimization, the reduced $\chitwo$ value with $N=3$ is minimized (see Eq.\,8 of \citetalias{Kunitomo+Guillot21}).
When we examine the quality of the solution, we also calculate $\chitwo_{N=6}$ by adding the surface helium abundance $\Ys$, the location of the convective-radiative boundary $\RCZ$, and root-mean-square sound speed rms($\delcs$) (see Fig.\,\ref{fig:A2}), or $\chitwo_{N=8}$ by further adding the surface Li abundance $\ALi$, and the surface Be abundance $\ABe$ (see Table\,\ref{tab:chi2-results-output}).

\section{Results} \label{sec:results}

\begin{figure}[!t]
  \begin{center}
    \includegraphics[angle=0,width=\hsize]{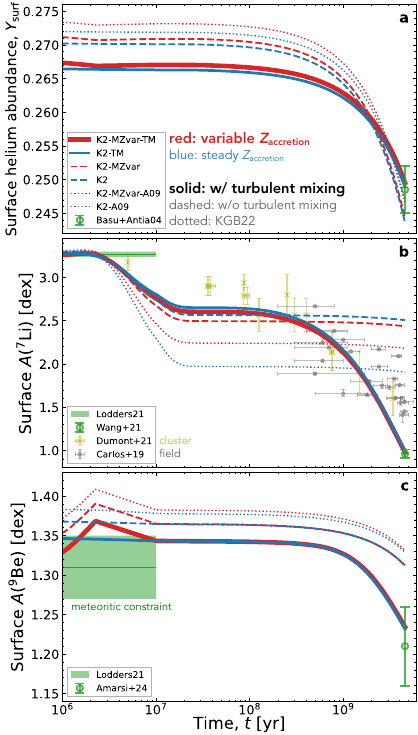}
  \end{center}
  \caption{
  Evolution of the surface abundances of helium (panel a), lithium (b), and beryllium (c). The red and blue lines show the models with variable and steady $\Zacc$, respectively. The solid and dashed lines show the models with and without turbulent mixing, respectively. The dotted lines show the models from \citetalias{Kunitomo+22} (with \citetalias{Asplund+09} abundances, without turbulent mixing but with overshooting). See Table\,\ref{tab:chi2} for more details.
  The green points show the observational constraints of the present-day Sun (see Table\,\ref{tab:targets}). The green shades show the meteoritic constraints \citep{Lodders21} arbitrarily extending from 1 to 10\,Myr. The crosses and small circles in panel b show the observed $\ALi$ values of clusters \citep[][]{Dumont+21} and individual stars \citep[][]{Carlos+19}, respectively, younger than the Sun.}
  \label{fig:t-HeLiBe}
\end{figure}

In this section, we show the evolutions of surface He, Li, and Be abundances, the metallicity profile, and neutrino fluxes, focusing on the models K2-MZvar-TM, K2-MZvar, K2-TM, and K2 (see Table\,\ref{tab:chi2}).

Figure\,\ref{fig:t-HeLiBe} shows the evolutions of $\Ys$, surface $\ALi$, and surface $\ABe$. The models with turbulent mixing (K2-TM and K2-MZvar-TM; solid lines) reproduce the three observations within $0.6\sigma$, as in \citet{Buldgen+25a}. The gradual depletion of the $\ALi$ values of solar-type stars \citep[][and references therein]{Carlos+19, Dumont+21} is also well reproduced by the models with the mixing.
In models without turbulent mixing (K2 and K2-MZvar; dashed lines), helium settling is too efficient and the observed $\Ys$ is not well reproduced ($1.4\sigma$ and $1.2\sigma$, respectively). Li and Be are not burned and thus their abundances are too high at the solar age ($\sim$30$\sigma$ and $\sim$2$\sigma$, respectively). In these models, the slight decreases of $\ALi$ and $\ABe$ are driven by gravitational settling, not nuclear burning.
Because He, Li, and Be depletion occurs in the MS, this behavior is independent of the protosolar accretion history.

Lithium is depleted by $\sim$0.6\,dex from 3 to 15\,Myr. This is because the temperature at the BCZ exceeds 2.5\,MK (see Fig.\,\ref{fig:schematic}a), and thus Li is burned \citep[but limited due to the short timescale; see also][]{Eggenberger+22, Buldgen+23}. The best models of \citetalias{Kunitomo+22} (dotted lines) deplete Li more (up to $\sim$1.3\,dex) in the pre-main sequence (pre-MS hereafter) because they include overshooting (see Appendix\,\ref{sec:ovs}).

We note the variety in the initial Li and Be abundances. In the calibration procedure (see Sect.\,\ref{sec:model}), the primordial metallicity $\Zproto$ is treated as a free parameter. Since we adopt the abundance scale of \citetalias{Asplund+21}, the initial Li and Be abundances are also adjusted accordingly during the calibration. Nevertheless, the initial $\ALi$ and $\ABe$ in the models with turbulent mixing agree with the meteoritic constraints provided by \citet{Lodders21}.

\begin{figure}[!t]
  \begin{center}
    \includegraphics[angle=0,width=\hsize]{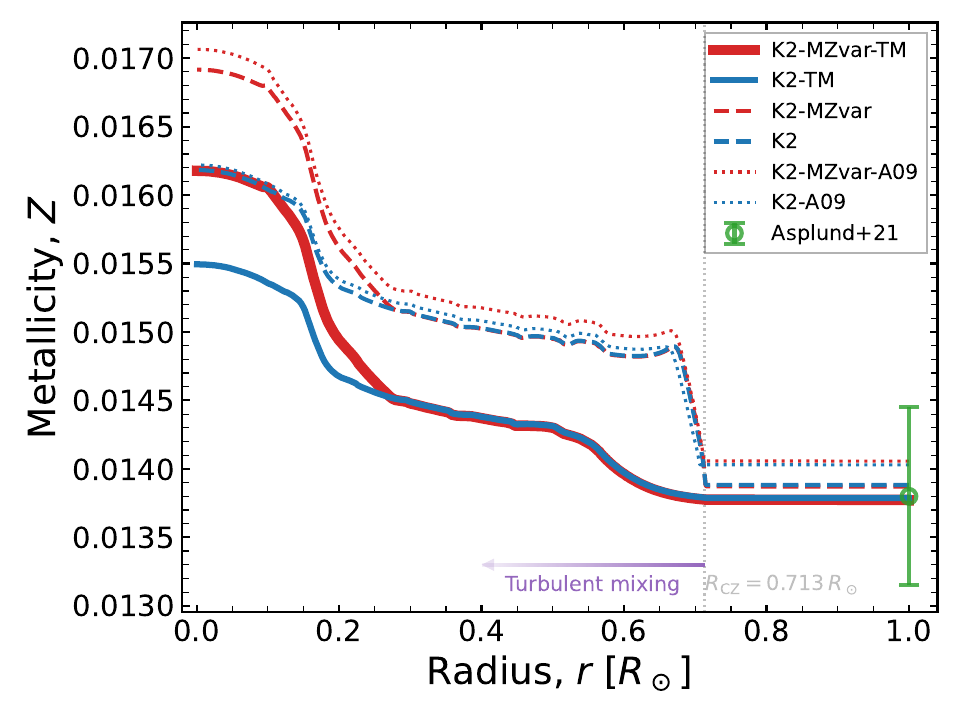}
  \end{center}
  \caption{
  Solar metallicity profile at the present day of the six models as in Fig.\,\ref{fig:t-HeLiBe}. The dashed vertical line shows the BCZ. See also Fig.\,\ref{fig:r-D} for the $\Dmix$ profile.}
  \label{fig:r-Z}
\end{figure}

\begin{figure}[!t]
  \begin{center}
    \includegraphics[angle=0,width=\hsize]{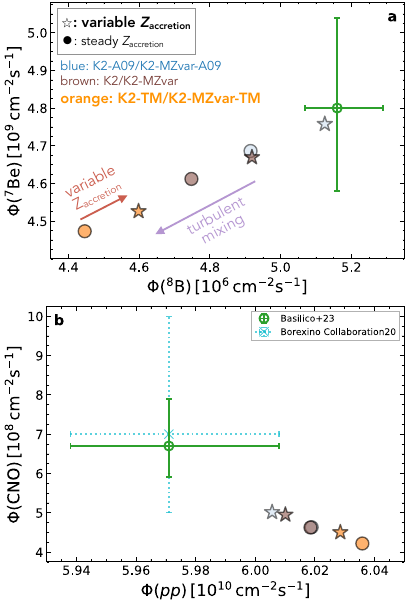}
  \end{center}
  \caption{Solar neutrino fluxes. The star and circle symbols show the models with variable and steady $\Zacc$, respectively. The orange and brown colors illustrate the models with and without turbulent mixing, respectively. The blue color shows the models from \citetalias{Kunitomo+22} (with \citetalias{Asplund+09} abundances without turbulent mixing).
  The points with error bars are the observational constraints (Table\,\ref{tab:targets}). The dotted, cyan error bar shows an old constraint by \citet{Borexino-Collaboration20}.}
  \label{fig:neu}
\end{figure}

Figure\,\ref{fig:r-Z} shows the present-day metallicity profile. The turbulent mixing is apparent below the BCZ: the metallicity jumps at the BCZ in the models without it because of gravitational settling, whereas turbulent mixing smooths the metallicity profile down to $\sim$0.5\,$\Rsun$. This corresponds to the extent from the BCZ ($=0.713\,\Rsun$) reached by diffusion by the solar age. The diffusion timescale $\tmix$ at a scale $\Delta r$ is estimated as
\begin{eqnarray}\label{eq:tmix}
    \tmix=\frac{\Delta r^2}{\Dmix} \sim  4.4\,\mathrm{Gyr} \left( \frac{\Delta r}{0.17\,\Rsun} \right)^2\left( \frac{\Dmix}{10^3\,\mathrm{cm^2s^{-1}}} \right)^{-1}\,.
\end{eqnarray}

The central metallicity, $\Zc$, is also lower because gravitational settling is suppressed. In models with variable and steady $\Zacc$, the turbulent mixing decreases $\Zc$ by 4.4\% and 4.3\%, respectively. The finding in \citetalias{Kunitomo+Guillot21}, that is, variable $\Zacc$ can increase $\Zc$ by up to $\sim$5\%, is still valid even for the models with turbulent mixing (4.4\%). Therefore, the effects of variable $\Zacc$ and turbulent mixing are counterbalanced: the model with turbulent mixing and variable $\Zacc$ has almost the same $\Zc$ as the model with steady $\Zacc$ without mixing.

The lower $\Zc$ due to turbulent mixing directly impacts neutrino fluxes. Figure\,\ref{fig:neu} shows that the new models with turbulent mixing (orange points) do not reproduce the observed neutrino fluxes.
Their $\phiB$ and $\phiBe$ are lower than for the models without turbulent mixing (brown) and for the observations. These fluxes are strongly sensitive to the central temperature $\Tc$ \citep[$\phiB\propto \Tc^{25}$ and $\phiBe\propto \Tc^{11}$;][]{Bahcall+Ulmer96}; thus, the lower fluxes come from the lower $\Tc$. There are two reasons for the lower $\Tc$: the lower $\Zc$ due to the fact that mixing reduces the central opacity and $\Tc$ as well. In addition, turbulent mixing suppresses helium settling, leading to a lower mean molecular weight at the center, $\muc$. For the hydrostatic equilibrium (pressure $\propto T\rho/\mu$), this leads to the lower $\Tc$.

Figure\,\ref{fig:neu}b shows that turbulent mixing reduces $\phiCNO$. This flux is also highly sensitive to $\Tc$ and, in addition, to $\Zc$. Both lower $\Tc$ and $\Zc$ of the models with mixing lead to a lower $\phiCNO$.
By contrast, their $\phipp$ is higher. This flux depends on $\Xc^2$, where $\Xc$ is the central hydrogen abundance. Because of inefficient helium settling, $\Xc$ is higher. We note that the lower $\Tc$ is counterbalanced by a higher $\Xc$ and thus the observed luminosity, $\Lsun$, is well reproduced.

As a result, the neutrino fluxes of model K2-TM (with mixing and steady $\Zacc$; orange circle in Fig.\,\ref{fig:neu}) are far from the observations: $\phiB$, $\phiBe$, $\phipp$, and $\phiCNO$ are within 
$7.9\sigma$,
$1.5\sigma$,
$1.8\sigma$, and
$3.1\sigma$.

\citetalias{Kunitomo+22} found that a higher $\Zc$ due to variable $\Zacc$ leads to higher neutrino fluxes, matching the observations.
The fluxes of our fiducial model with turbulent mixing with variable $\Zacc$ (K2-MZvar-TM; orange star) are indeed closer to observations compared to K2-TM. Thus, the fact that a higher $\Zc$ due to variable $\Zacc$ leads to higher neutrino fluxes remains valid in the models with turbulent mixing.
However, the effect of accretion is not sufficient: model K2-MZvar-TM is distant from the observations by
6.2$\sigma$ for $\phiB$,
1.2$\sigma$ for $\phiBe$,
1.6$\sigma$ for $\phipp$, and
2.7$\sigma$ for $\phiCNO$.
In particular, $\phiB$ and $\phiCNO$ still poorly match observations.
Therefore, we conclude that, although variable $\Zacc$ is still a key process to explain the neutrino fluxes, we need another physical process (Sect.\,\ref{sec:discussions}) to reproduce at the same time the light element abundances, $\ALisun$ and $\ABesun$, and the neutrino fluxes.

Here we explain why the models in \citetalias{Kunitomo+22} (blue) have higher $\phiB$ and $\phiBe$ than the new models without mixing (brown) even though these models have similar $\Zc$ (Fig.\,\ref{fig:r-Z}).
This results from a higher $\Tc$ in the \citetalias{Kunitomo+22} models, which is caused by a higher $\muc$. The increase in $\muc$ originates from a higher central helium abundance $\Yc$, which in turn results from the extended calibration: \citetalias{Kunitomo+22} adopted six input parameters ($N=6$), including $\Ys$. Since the \citetalias{Kunitomo+22} models do not include turbulent mixing, gravitational helium settling is efficient. To match the observed $\Ys$, the initial abundance $\Yproto$ has to be higher (see Fig.\,\ref{fig:t-HeLiBe}a), and thus the bulk $Y$ and $\Yc$ are also higher.
From the comparison with the models with $N=3$ and \citetalias{Asplund+09} abundances, we confirmed this calibration effect: these models with steady or variable $\Zacc$ have slightly lower $\Yc$, $\phiB$, and $\phiBe$ values than models K2 or K2-MZvar.

\section{Discussions} \label{sec:discussions}

\subsection{Origin of turbulent mixing}
\label{sec:TM}

The presence of turbulent mixing at the BCZ has already been mentioned in the 1990s as missing from the SSM \citep{Christensen-Dalsgaard+96}. Its exact physical origin is yet unknown, as the BCZ is the place of multiple physical phenomena that could lead to additional mixing of chemicals. \citet{Brun+02} link this mixing to the presence of the solar tachocline, which coincides with the BCZ. More recently, \citet[][]{Eggenberger+22} constructed solar models including circulation, shear instability, and magnetic Tayler instability. These models could be calibrated to reproduce the Li depletion observed in the Sun. However, the revision of the Be abundance \citep{Amarsi+24} showed that only a small amount of Be was burned at the solar age, implying a shallow depth of turbulent mixing at the BCZ. This was shown to be in disagreement with models including the effects of rotation \citep{Buldgen+24b, Buldgen+25a}, as models including circulation, shear instability, and magnetic Tayler instability present an extended mixed region down to $\sim$0.4\,$\Rsun$, while observations of Be forbid a mixing below $\sim$0.6\,$\Rsun$. It is unclear yet whether the depletion of Li and Be is directly linked to the flat rotation profile in the solar radiative interior \citep[down to $\sim$0.2\,$\Rsun$; ][]{Couvidat+03} and there is no consensus on the underlying physical mechanism responsible for angular momentum transport in the solar interior.

One might think that overshooting is the origin of the mixing below the BCZ. \citetalias{Kunitomo+22} adopted diffusive overshooting, which enhances Li depletion in the pre-MS phase (Sect.\,\ref{sec:results}). Indeed, with a more efficient diffusive overshooting, $\ALisun$ can be reproduced (Fig.\,\ref{fig:fov}b). However, the observed gradual decrease of the surface $\ALi$ of solar-type stars cannot be reproduced, and a slight decrease of $\ABe$ from the proto-solar phase to the present-day, which observations suggest, is also not made. We will discuss the effect of diffusive overshooting in Appendix\,\ref{sec:ovs} in more detail.

Although the diffusive overshooting model \citep{Herwig00} is widely used, other overshooting models have also been discussed.
\citet{Zhang+19} suggested that the inclusion of an overshooting model could solve the sound speed discrepancies and the Li depletion. However, while the sound speed can be corrected, the model of \citet{Zhang+19} does not allow us to correct the density profile in the convective envelope (see model OV09Ne in their Fig.\,1). Recent works by \citet{Baraffe+21} and \citet{Baraffe+22} have shown that convective penetration is not expected to fully correct the sound speed anomaly and extend deep enough to burn both Li and Be based on multi-dimensional hydrodynamic simulations. 
\citet{Zhang+23} have suggested another overshooting model and tested it on asteroseismic observations of a B-type star. The application of this new model to the case of the base of the solar convective envelope could have an impact on the depletion of Li and Be.

\subsection{Solar evolutionary models with rotation}

In this study, we considered turbulent mixing below the BCZ \citep[][see Sect.\,\ref{sec:model}]{Proffitt+Michaud91,Buldgen+25a}.
Although this is likely to be related to rotational mixing, the exact underlying process is still under debate (Sect.\,\ref{sec:TM}). In this study, we did not solve angular momentum evolution and instead adopted an empirical mixing model in non-rotating solar models. In addition to the physical origin of the mixing, there is also a practical numerical issue in the treatment of meridional circulation.
In principle, the circulation should be calculated using an advective term in the angular momentum transport equation, and some stellar evolution codes follow this implementation, such as GENEC \citep[][]{Eggenberger+22}, Cesam2k20 \citep[][Manchon et al., in preparation]{Marques+13}, and STAREVOL \citep[][]{Palacios+06, Decressin+09}.
However, due to numerical difficulties, others, such as MESA \citep{Paxton+13}, PARSEC \citep[][]{Nguyen+22} and YREC \citep{Yang+25}, treat circulation as a diffusion term \citep[see discussions in][]{Potter+12, Salaris+Cassisi17}. 
In the future, rotating solar models should be developed and reproduce self-consistently all the solar observations \citep[i.e., $\ZXs$, $\cs$ profile, neutrino fluxes, Li and Be abundances, and rotation profile;][]{Eggenberger+22}, as well as be consistent with asteroseismic constraints on the internal rotation \citep[e.g.,][]{Buldgen+Eggenberger23, Dumont23}.

Regarding rotational mixing, the initial rotation rate is a key parameter. The protoplanetary disk can regulate the rotation period of pre-MS stars via star-disk interaction \citep[so-called ``disk-locking''; see, e.g.,][]{Amard+Matt23, Takasao+25}. The longer disk lifetime leads to a slow rotator, which may have a stronger shear in the interior and thus a more efficient Li depletion \citep[][]{Eggenberger+12}. Also, the long disk lifetime is likely to lead to a higher chance of giant planet formation. \citet{Bouvier08} suggested a possible link between Li abundance and exoplanet occurrence rate. This link should be investigated in more detail in future work.

\subsection{Li depletion by cold accretion}

We note the effect of protostellar accretion on Li depletion. Accretion affects the thermal evolution of protostars \citep[e.g.,][]{Hartmann+97, Kunitomo+17}. \citet{BC10} found that if the accretion entropy is quite low (so-called cold accretion scenario), the stellar interior is hot enough to quickly deplete Li in the accretion phase \citep[see also][]{Tognelli+20}. However, \citet{Kunitomo+17} concluded from a comparison with young clusters on the Hertzsprung--Russell diagram that most stars should not have experienced cold accretion. Furthermore, again, this scenario does not explain the observed gradual decrease in $\ALi$ of solar-type stars.

\subsection{A new protosolar helium abundance}

Figure\,\ref{fig:t-HeLiBe}a and Table\,\ref{tab:chi2-results-input} show that the models with turbulent mixing that reproduce the constraints on the observed atmospheric abundances of lithium and beryllium have a proto-solar helium abundance ($\Yproto$) ranging from 0.2648 to 0.2659.
Reporting the uncertainty on the present-day solar atmospheric helium abundance from \citet{Basu+Antia04} to the values above, we find a revised value $\Yproto=0.2654 \pm 0.0035$.
This is lower than the value derived from classical evolution models of the Sun, $\Yproto\sim0.278\pm0.006$ \citep[][]{Serenelli+Basu10}. 

\begin{figure}[!t]
  \begin{center}
    \includegraphics[angle=0,width=\hsize]{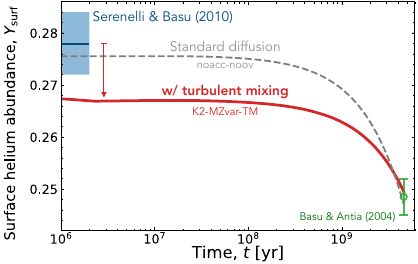}
  \end{center}
  \caption{
  Evolution of $\Ys$ of model K2-MZvar-TM (red solid line) and a non-accreting model from \citetalias{Kunitomo+Guillot21} with standard atomic diffusion (i.e., no overshooting and turbulent mixing; gray dashed) starting with a high $\Yproto$ value within the range suggested by \citet[][blue shaded region]{Serenelli+Basu10}.
  Turbulent mixing suppresses gravitational settling and thus leads to a reduced $\Yproto$ value.}
  \label{fig:Yproto}
\end{figure}

Figure\,\ref{fig:Yproto} demonstrates that turbulent mixing lowers the estimated $\Yproto$ value. Turbulent mixing suppresses gravitational helium settling, and thus model K2-MZvar-TM has a low $\Yproto$ value and successfully reproduces the present-day $\Ys$ constraint. Although the non-accreting model ``noacc-noov'' from \citetalias{Kunitomo+Guillot21} that includes standard atomic diffusion (i.e., no overshooting and turbulent mixing) also reproduces $\Ys$, this model has a high $\Yproto$ value\footnote{The high $\Yproto$ value of ``noacc-noov'' is derived by extended calibration that includes $\Ys$ in target quantities (see tables 1, A.1, and A.2 of \citetalias{Kunitomo+Guillot21}).} within the range suggested by \citet{Serenelli+Basu10} and causes efficient helium settling.

While the inclusion of additional mixing at the BCZ leads to an incompatibility with neutrino fluxes (Sect.\,\ref{sec:neu}), they do reproduce the surface He, Li and Be abundances at the age of the Sun better than SSMs (see Fig.\,\ref{fig:t-HeLiBe} and Table\,\ref{tab:chi2-results-output}). Therefore, it appears that our description of the efficiency of mixing at the BCZ is more realistic than that of SSMs. This implies that the value we provide for $\Yproto$ can be considered quite reliable but should be revised after any potential future improvements to the physical ingredients of solar models.

The constraint on $\Yproto$ from solar modeling is a key ingredient for interior models of Jupiter and Saturn \citep{Guillot05}: The precise determination of the masses and shapes of these planets (and therefore their mean density) implies that the bulk abundance of heavy elements (all elements other than hydrogen and helium) that is inferred to be present in the interiors of these planets is inversely correlated to $\Yproto$. Recently, the precise measurements of the masses and gravitational moments by Juno in Jupiter \citep{Iess+2018} and Cassini in Saturn \citep{Iess+2019} have led to a decisive tightening of the constraints on interior models. This has resulted in tensions between a very low value of the enrichment in heavy elements in the outer envelopes of these planets, as favored by interior models, compared to atmospheric constraints from spectroscopy and in situ measurements that yield clear enrichments over solar values \citep[see][and references therein]{Mankovich+Fuller2021, Howard+2023, Guillot+23}. 

Models of Jupiter and Saturn \citep[e.g.,][]{Guillot+18} have so far relied on the constraint derived from classical evolution models of the Sun ($\Yproto\sim0.278\pm0.006$; see above), while our new value is lower by 0.0126, equivalent to a global enrichment in heavy elements by approximately one-time solar value. Using this revised value may help ease some of the tension between interior models and spectroscopy, although for Jupiter this effect will be limited by the fact that the helium abundance in the planet's atmosphere has been precisely measured by the Galileo probe \citep{vonZahn+1998}.

\subsection{Explaining the neutrino fluxes}\label{sec:neu}

Our models with turbulent mixing have lower $\phiB$, $\phiBe$, and $\phiCNO$ than the observed constraints, even in the case with variable $\Zacc$ (Fig.\,\ref{fig:neu}). We discuss hereafter several effects that might account for this disagreement: input physics (nuclear reactions, opacity, solar winds, and variable $\Zacc$ model) and observations' uncertainties (particularly on CNO neutrinos).

Neutrino fluxes can be influenced by the ingredients of solar models, including the formalism used for electron screening, nuclear reaction rates, and opacities. 
Electron screening seems to be the most promising avenue; the current formalism applied in all stellar evolution codes is that of static screening, which neglects the velocity differences between electrons and ions. Recently, \citet[][]{Dappen24} wrote a brief review of the current state of the issue, highlighting that this effect still needs to be studied in detail as it impacts core conditions and neutrino fluxes, potentially leading to higher central temperature and metallicity in calibrated solar models.
Updated nuclear reaction rates have been recently published \citep[Solar Fusion III; ][]{Acharya+24} and could slightly modify the values found by calibrated solar models. Regarding opacities, the physical origin of the observed differences between theoretical computations and the measurement of \citet{Bailey+15} and \citet{Nagayama+19} is still unknown, and measurements at higher temperatures and electron densities have not been achieved yet. Therefore, the measured differences cannot be reliably extrapolated to conditions deeper in the Sun, implying that the expected impact on the solar core conditions and the neutrino fluxes is yet unknown. 
All three effects remain potential progress avenues that would alter the comparisons between solar models and neutrino experiments.

\citet[][see their Fig.\,18]{Zhang+19} showed that solar winds can increase $\Zc$ \citep[see also][]{Sackmann+Boothroyd03,Wood+18}. Both observational and theoretical studies have suggested that the young Sun had more vigorous winds than presently \citep[][]{Wood+05, Suzuki+13}. If solar winds are metal-rich (e.g., first ionization potential effect), then the metallicity gradient becomes larger with time. This scenario is interesting, but the history of mass-loss rate and composition evolution is still quite uncertain. This needs to be investigated in more detail in the future \citep[][]{Buldgen+25c}.

The variable $\Zacc$ increases $\phiB$, $\phiBe$, and $\phiCNO$. In this study, we used the same $\Zacc$ model as in the best model of \citetalias{Kunitomo+22} (i.e., $M_1=0.90\,\Msun$, $M_2=0.96\,\Msun$, and $\Zaccmax=0.065$). This is determined by complex dust dynamics in the protosolar disk. Moreover, we fixed the abundance scale to \citetalias{Asplund+21} and changed the metallicity, but each element can behave differently. If we expand the model further to treat the abundance of each element in the accretion flow, then the central CNO abundance (and thus CNO neutrino fluxes) may be enhanced. An interdisciplinary study integrating knowledge from solar/stellar physics, planet formation theory, disk chemistry, and Solar-System studies will be needed in the future.

Finally, we note that although truly substantial efforts have been made on the observational side, the interpretation of the observed neutrino fluxes is still challenging. In particular, CNO neutrinos have been detected only in the Borexino experiment. We also note that the $\phiCNO$ constraint has been updated: the new constraint \citep[green circle in Fig.\,\ref{fig:neu}b;][]{Basilico+23} is slightly lower than the previous one \citep[cyan dotted error bar; ][$\phiCNO=(7.0^{+3.0}_{-2.0})\times10^8\,\mathrm{cm^{-2}s^{-1}}$]{Borexino-Collaboration20}, but has a much smaller uncertainty. Both constraints were derived from the data of the Borexino experiments, but different analysis techniques were used. The best model in \citetalias{Kunitomo+22} agreed with the old constraint by 0.99$\sigma$, while the new constraint leads to a worse value of 2.11$\sigma$. The best model in this paper, K2-MZvar-TM, has 1.2$\sigma$ and 2.7$\sigma$, respectively. Given the direct impact of a revision of the precision of the observational constraints on the conclusions drawn on solar models, an independent measurement of $\phiCNO$ would be a strong confirmation of the current observed disagreements.

\section{Conclusion} \label{sec:conclusion}

The structure and evolution of the Sun have usually been tested by comparing with observed surface metallicity, helioseismic constraints, and neutrino fluxes. Our previous study, \citetalias{Kunitomo+22}, showed that protosolar accretion can lead to an increased central metallicity by up to $\approx$5\%, thus significantly improving the agreement between theoretical and measured neutrino fluxes.
Since this study, recent refinements of the lithium and beryllium abundances on the solar surface have highlighted the importance of turbulent mixing in the solar interior. This led us to develop solar models that include both protosolar accretion and turbulent mixing to account for the measured surface Li and Be abundances.

Our models with turbulent mixing reproduce well the observed He, Li, and Be abundances within $0.3\sigma$, $0.6\sigma$, and $0.5\sigma$, respectively, consistent with previous studies \citep[][]{Buldgen+25a}. This is a significant improvement from models without mixing, which reproduce them within $\sim$1.4$\sigma$, $\sim$30$\sigma$, and $\sim$2.0$\sigma$, respectively.
The difference comes from two effects of the mixing: it promotes nuclear burning to deplete Li and Be and suppresses the gravitational settling of He.
The models suggest a lower protosolar He abundance, $\Yproto=0.2651 \pm 0.0035$, than previously suggested.

The limited settling due to turbulent mixing has a negative impact on the central metallicity $\Zc$, which is decreased by $\approx$4.4\%.
This is almost the same amount as that of the increase ($\approx$4.4\%) due to protosolar accretion with variable composition, as found by \citetalias{Kunitomo+Guillot21}.
This leads to neutrino fluxes in disagreement with observations: for the model K2-TM with turbulent mixing and steady $\Zacc$, 
by $7.9\sigma$, $1.5\sigma$, $1.8\sigma$, and $3.1\sigma$ for $\phiB$, $\phiBe$, $\phipp$, and $\phiCNO$.
For the model K2-MZvar-TM with turbulent mixing and variable $\Zacc$, the situation improves but remains insufficient: 6.2$\sigma$, 1.2$\sigma$, 1.6$\sigma$, and 2.7$\sigma$, respectively. Therefore, while variable $\Zacc$ remains an important effect to reproduce the observed neutrino fluxes, as it has a strong impact on the central metallicity, it is insufficient on its own.

Further investigations of the physical mechanisms responsible for internal mixing in the Sun are needed to understand the magnitude of turbulent mixing and its consequences for the Sun's internal structure and evolution. Although they are likely to be related to rotation, \citet{Buldgen+25a} suggested the need for an improvement from the state-of-the-art model in \citet{Eggenberger+22}. In this study, we did not consider rotation, but instead adopted an empirical turbulence model treated as a diffusion coefficient related to density as in \citet{Proffitt+Michaud91, Buldgen+25a}, to model the transport of chemicals. The investigation of the underlying physics of the light element depletion observed in the Sun should ideally be coupled to the derivation of a self-consistent and physically motivated description of the evolution of angular momentum. The remaining differences of neutrino fluxes between observations and our models that include both turbulent mixing and protosolar accretion likely require further examination of input physics such as electron screening, nuclear reaction rates, the impact of solar winds, and the $\Zacc$ model itself \citep[][]{Buldgen+25c}. Furthermore, a higher precision on the determination of neutrino fluxes, especially $\phiB$ and $\phiCNO$, would prove essential in further constraining the physical ingredients of solar models.

\section*{Data availability}
Tables\,\ref{tab:chi2-results-input} and \ref{tab:chi2-results-output} and supplemental materials are available on Zenodo at \url{https://doi.org/10.5281/zenodo.16789192}.

\begin{acknowledgements}
    We are grateful to Ebraheem Farag for his valuable comments on MESA. M.K. thanks Observatoire de la C\^{o}te d'Azur for the hospitality during his long-term stay in Nice. This work was supported by JSPS KAKENHI Grant Nos. JP23K25923, JP24K07099, and JP24K00654. G.B. acknowledges fundings from the Fonds National de la Recherche Scientifique (FNRS) as a postdoctoral researcher. T.G. acknowledges funding from the {\it Programme National de Plan\'{e}tologie}. Numerical computations were carried out on the PC cluster at the Center for Computational Astrophysics, National Astronomical Observatory of Japan. Software: \texttt{MESA} \citep[version 12115;][]{Paxton+11,Paxton+13,Paxton+15,Paxton+18, Paxton+19}.
\end{acknowledgements}

\bibliographystyle{aa}

\begin{appendix}
\section{Effect of abundance update}\label{sec:abundance}

In this study, we used the solar surface abundances in \citetalias{Asplund+21}, while we adopted \citetalias{Asplund+09} in \citetalias{Kunitomo+Guillot21} and \citetalias{Kunitomo+22}. The $\ZXs$ value slightly increased mainly due to a higher neon abundance. We note that although \citetalias{Asplund+21} slightly updated $\Ys$ considering the uncertainties in the equation of state, we adopted the canonical value by \citet{Basu+Antia04}.
The assumed abundances have four effects: target abundances in optimization (see Table\,\ref{tab:targets}), opacity tables, a different initial seed structure, and required opacity increase.

We used the OPAL opacity tables for high-temperature regions obtained from the OPAL website with the chemical mixture of \citetalias{Asplund+09} with the updated neon abundance in \citet[][]{Young18}, which is essentially the same as \citetalias{Asplund+21}.
We used the \citet{Ferguson+05} opacities with \citetalias{Asplund+21} abundances for low-temperature regions, which are available in the MESA version 24.08.1.
We note that OPLIB opacity tables with \citetalias{Asplund+21} are also available in the latest version of MESA \citep[][]{Farag+24}. We chose OPAL in this study because OPLIB opacities are lower than OPAL ones by $\sim$10--15\% in the high-temperature ($\sim$$10^7$\,K) region (likely due to the effect of different equations of state and electron density, but still under debate). \citet[][see their Fig.\,5]{Salmon+21} and \citet[][see their Table\,2]{Buldgen+24b} showed that SSMs with OPLIB have much lower $\phiB$, $\phiBe$, and $\phiCNO$ than those with OPAL.

We created an initial seed with mass $0.1\,\Msun$, radius $4\,\Rsun$, and metallicity $0.02$ with the \citetalias{Asplund+21} abundance and the opacities described above.
We used the photospheric metal abundance scale of \citetalias{Asplund+21} (see their table 2) for the seed and accreted materials.

\begin{figure}[!t]
  \begin{center}
    \includegraphics[angle=0,width=\hsize]{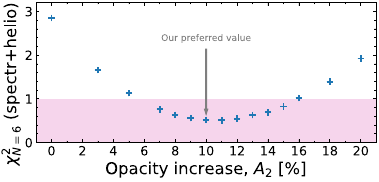}
    \includegraphics[angle=0,width=\hsize]{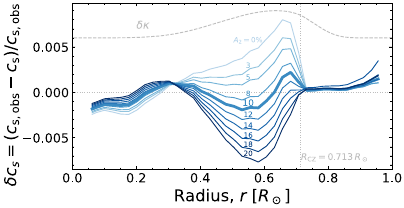}
  \end{center}
  \caption{Optimized results of the models ``K2-A2'' (see Table\,\ref{tab:chi2}) with opacity increase (see Eqs.\,\ref{eq:kap} and \ref{eq:delkap}). Top: $\chitwo$ values with $N=6$ as a function of the opacity increase parameter, $A_2$.
  Bottom: Sound speed profile of models with $A_2=0$, 3, 5, 8, 10, 12, 14, 16, 18, and 20\% at the solar age, compared with helioseismic data in \citet{Basu+09}, $\csobs$. The model with our preferred value, $A_2=10\%$, is highlighted with the thick line. The dashed line illustrates the shape of the $\delta_\kappa$ function with an arbitrary $+0.006$ vertical shift.}
  \label{fig:A2}
\end{figure}

Finally, opacity increase is widely considered in modern solar models to reconcile the sound speed profile \citep[][]{Christensen-Dalsgaard+09, Serenelli+09, Bailey+15, Buldgen+19, Buldgen+25a}. We showed in \citetalias{Kunitomo+Guillot21} that opacity increase with a parameter $A_2$ by 12\%--18\% significantly improves the fit to the observational $\cs$ profile in the case with the \citetalias{Asplund+09} abundances and took 12\% as a fiducial value in \citetalias{Kunitomo+22}. We performed a parameter study to derive the required opacity increase in the case with the slightly more metal-rich \citetalias{Asplund+21} abundances and the corresponding OPAL opacity tables (model ``K2-A2''; see Table\,\ref{tab:chi2}). Here we adopted the SSM framework (i.e., optimizing $\Yproto$, $\Zproto$, and $\amlt$ to match observed $\Teff$, $\Lstar$, and $\ZXs$) and included neither turbulent mixing nor overshooting. Accretion with a constant $\Zacc$ ($=\Zproto$) is considered.
Figure\,\ref{fig:A2} shows that the $\chitwo_{N=6}$ value ($N=6$)\footnote{For optimization, $N=3$ as described above, but here we also include $\Ys$, $\RCZ$, and rms($\delcs$) values to see the quality of the solutions.}
is minimized with $A_2\simeq10\%$--$12\%$, which is slightly lower than the results with \citetalias{Asplund+09} described above but in agreement with a recent helioseismic constraint by \citet{Buldgen+25b}. Therefore, we took $A_2=10\%$ throughout this paper.

\begin{figure}[!t]
  \begin{center}
    \includegraphics[angle=0,width=\hsize]{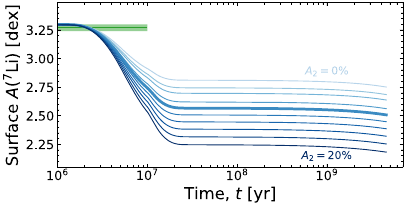}
  \end{center}
  \caption{Evolution of the surface lithium abundance, $\ALi$, of the models ``K2-A2''. The lines are color-coded by $A_2$ as in the bottom panel of Fig.\,\ref{fig:A2}. We note that the observational constraint of the present-day Sun, $\ALisun=0.96\pm0.05$\,dex, is not displayed in this plot.}
  \label{fig:A2_Li}
\end{figure}

Figure\,\ref{fig:A2_Li} shows that a higher $A_2$ value leads to a stronger Li depletion in the pre-MS (from $\sim$3 to $\sim$20\,Myr).
This is because higher opacity at around the BCZ shifts $\RCZ$ downward and thus increases the temperature at the BCZ. Although this effect changes $\ALi$ by up to $\sim$0.5\,dex, the observational constraint of the present-day Sun, $\ALisun=0.96\pm0.05$\,dex, is even much lower than the extreme model with $A_2=20\%$, which leads to $\ALi\sim2.2$\,dex.

\section{Models with diffusive overshooting}\label{sec:ovs}

Here we show models ``K2-fov'' including diffusive overshooting, which can reproduce the solar Li abundance but are inconsistent with the observed trend of other solar-type stars.
Diffusive overshooting corresponds to the model where the diffusion coefficient $\Dmix$ exponentially drops with the depth from the BCZ as \citep{Herwig00}
\begin{eqnarray}\label{eq:Dov}
    \Dmix=D_0\exp \left(\frac{-2\Delta r}{\fov H_P} \right)\,,
\end{eqnarray}
where $D_0$ is the convective diffusion coefficient at the BCZ, $\Delta r$ is the distance from the BCZ, $H_P$ is the pressure scale height at the BCZ, and $\fov$ is a dimensionless parameter.

\begin{figure}[!t]
  \begin{center}
    \includegraphics[angle=0,width=\hsize]{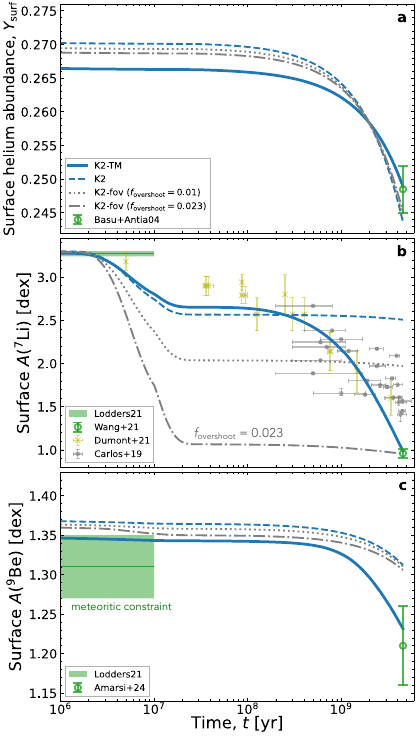}
  \end{center}
  \caption{ Same as Fig.\,\ref{fig:t-HeLiBe} but showing models ``K2-fov'' including diffusive overshooting with $\fov=0.023$ (gray dot-dashed line) and $\fov=0.01$ (dotted).}
  \label{fig:fov}
\end{figure}

\begin{figure}[!t]
  \begin{center}
    \includegraphics[angle=0,width=\hsize]{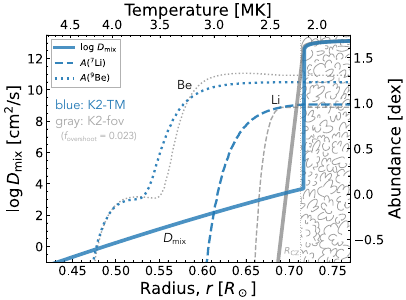}
  \end{center}
  \caption{
  Profiles of $\Dmix$ (solid lines) of models ``K2-TM'' (blue) and ``K2-fov'' with $\fov=0.023$ (gray) at the solar age. The dashed and dotted lines show the profiles of Li and Be abundances, respectively. The thin vertical dotted line shows the location of $\RCZ=0.713\,\Rsun$.}
  \label{fig:r-D}
\end{figure}

Figure\,\ref{fig:fov} shows the evolution of the surface He, Li, and Be abundances, as in Fig.\,\ref{fig:t-HeLiBe}. As described in Sect.\,\ref{sec:results}, even the models without turbulent mixing or overshooting deplete Li in the pre-MS. Overshooting makes this Li burning more efficient by effectively deepening the surface CZ \citep[see also][]{Eggenberger+22, Buldgen+23}. The model with $\fov=0.023$ can reproduce $\ALisun$. However, in the MS, the Li burning is negligible, and thus $\ALi$ is almost constant, unlike the observation of solar-type stars.
Figure\,\ref{fig:r-D} shows the $\Dmix$ profile. The $\Dmix$ of overshooting (Eq.\,\ref{eq:Dov}) rapidly drops just below the BCZ, whereas turbulent mixing exhibits a long tail down to $\sim$0.5\,$\Rsun$.

In addition to the constant $\ALi$ evolution in the MS, the shallow overshooting mixing has other problems: Figure\,\ref{fig:fov}a shows that helium settling is not suppressed and thus the observed $\Ys$ is not reproduced. Overshooting does not reach the Be-burning region ($\sim$3.5\,MK) and thus $\ABe$ is not sufficiently depleted (Fig.\,\ref{fig:fov}c; the slight depletion after $\sim$1\,Gyr is due to settling).

From the results above, we conclude that models with diffusive overshooting alone do not satisfy all the observed constraints simultaneously and suggest a deeper mixing in the radiative region.

\section{Details of simulation results}\label{sec:tables}

In this Appendix, we provide the details of our simulation results. Tables \ref{tab:chi2-results-input} and \ref{tab:chi2-results-output} show the input parameters and results at 4.567 Gyr, respectively, which were optimized using the simplex method (Sect.\,\ref{sec:model}).

Additional supplemental materials are available on \href{https:doi.org/10.5281/zenodo.16789192}{Zenodo}. These include a csv file summarizing the optimized input parameters and results of all the simulation models, and structure and evolution data of the optimized cases of models K2-MZvar-TM, K2-TM, K2-MZvar, and K2. In addition, animations of the evolutions of the optimized case of each model are also available on Zenodo.
We used the input files of the MESA code provided in \citetalias{Kunitomo+Guillot21}\footnote{\url{https://zenodo.org/records/5506424}}, with modifications for turbulent mixing, \citetalias{Asplund+21} abundance scale, and opacity tables (see details in Appendix\,\ref{sec:abundance}).

\onecolumn

\tiny

\begin{longtable}{l|llllllllllll}
        \caption{Input parameters minimized using chi-squared simulations. \label{tab:chi2-results-input}}\\
 	\hline \hline
            Model name & $\amlt$ & $\fov$ & $n$ & $\DT$ & $A_2$ & $\Xaccini$ & $\Yaccini$ & $\Zaccini$ & \tablefootmark{(a)}   \\
            Unit & & & & [$\mathrm{cm^2s^{-1}}$] & & & & \\
            \hline
            \endfirsthead
            \caption{continued.}\\
            \hline\hline
            Model name & $\amlt$ & $\fov$ & $n$ & $\DT$ & $A_2$ & $\Xaccini$ & $\Yaccini$ & $\Zaccini$ & \tablefootmark{a}   \\
             & & & & [$\mathrm{cm^2s^{-1}}$] & & & & \\
            \hline
            \endhead
            \hline
            \endfoot
K2 & 1.812 & --- & --- & --- & 0.10 & 0.7163 & 0.2691 & 0.0147 & 1 \\
K2-MZvar & 1.806 & --- & --- & --- & 0.10 & 0.7159 & 0.2702 & 0.0139 & 2 \\
K2-TM & 1.787 & --- & 4 & 5000 & 0.10 & 0.7212 & 0.2648 & 0.0139 & 1 \\
{\bf K2-MZvar-TM} & 1.781 & --- & 4 & 5000 & 0.10 & 0.7210 & 0.2659 & 0.0131 & 2 \\
K2-fov10 & 1.808 & 0.010 & --- & --- & 0.10 & 0.7173 & 0.2682 & 0.0145 & 1 \\
K2-fov23 & 1.804 & 0.023 & --- & --- & 0.10 & 0.7181 & 0.2675 & 0.0144 & 1 \\
K2-A09 & 1.821 & 0.010 & --- & --- & 0.12 & 0.7143 & 0.2710 & 0.0147 & 1 \\
K2-MZvar-A09 & 1.817 & 0.004 & --- & --- & 0.12 & 0.7133 & 0.2727 & 0.0140 & 2 \\
         \end{longtable}
         \tablefoot{
         See Table\,\ref{tab:chi2} for the parameter settings of each model. This table is available in electronic form on \href{https:doi.org/10.5281/zenodo.16789192}{Zenodo}.
         \tablefoottext{a}{
         1: accretion with a steady composition, 2: accretion with a variable composition ($M_1=0.90\,\Msun$, $M_2=0.96\,\Msun$, and $\Zaccmax=0.065$).}}
         
\begin{longtable}{l|lllllll|llllll}
	\caption{Results minimized by the chi-squared simulations.
        \label{tab:chi2-results-output}}\\
    \hline\hline
            Model name & $\chi^2_{N=8}$ & rms($\delcs$) & $\ZXs$ & $\Ys$ & $\RCZ$ & $\ALi$ & $\ABe$ & $\Zc$ & $\phipp$ & $\phipep$ & $\phiBe$ & $\phiB$ & $\phiCNO$ \\
            Unit &  & [\%] & & & [$\Rsun$] & [dex] & [dex] & & [$10^{10}$]& [$10^{8}$]& [$10^{9}$]& [$10^{6}$]& [$10^{8}$] \\
            \hline
            \endfirsthead
            \caption{continued.}\\
            \hline\hline
            Model name & $\chi^2$ & rms($\delcs$) & $\ZXs$ & $\Ys$ & $\RCZ$ & $\Zc$ & $\ALi$ & $\ABe$ & $\phipp$ & $\phipep$ & $\phiBe$ & $\phiB$ & $\phiCNO$ \\
            Unit &  & [\%] & & & [$\Rsun$] & [dex] & [dex] & & [$10^{10}$]& [$10^{8}$]& [$10^{9}$]& [$10^{6}$]& [$10^{8}$] \\
            \hline
            \endhead
            \hline
            \endfoot
K2 & 121 & 0.107 & {\bf{0.01870}} &0.244 & {\bf{0.716}} &2.507 & 1.312 & 0.0162 & 6.019 & {\bf{1.444}} &{\bf{4.612}} &4.748 & 4.626  \\
K2-MZvar & 110 & 0.109 & {\bf{0.01870}} &0.244 & {\bf{0.716}} &2.436 & 1.312 & 0.0169 & 6.010 & {\bf{1.435}} &{\bf{4.669}} &4.919 & 4.950  \\
K2-TM & {\bf{0.191}} &{\bf{0.092}} &{\bf{0.01870}} &{\bf{0.249}} &{\bf{0.717}} &{\bf{0.986}} &{\bf{1.232}} &0.0155 & 6.036 & {\bf{1.453}} &4.473 & 4.446 & 4.219  \\
{\bf K2-MZvar-TM} & {\bf{0.209}} &0.102 & {\bf{0.01870}} &{\bf{0.250}} &{\bf{0.717}} &{\bf{0.976}} &{\bf{1.234}} &0.0162 & 6.029 & {\bf{1.445}} &4.526 & 4.598 & 4.502  \\
K2-fov10 & 51.7 & {\bf{0.100}} &{\bf{0.01870}} &0.244 & {\bf{0.715}} &1.969 & 1.310 & 0.0160 & 6.022 & {\bf{1.446}} &{\bf{4.583}} &4.686 & 4.539  \\
K2-fov23 & {\bf{0.68}} &{\bf{0.091}} &{\bf{0.01870}} &{\bf{0.245}} &{\bf{0.715}} &{\bf{0.955}} &1.306 & 0.0159 & 6.025 & {\bf{1.447}} &4.557 & 4.638 & 4.466  \\
K2-A09 & 45.8 & {\bf{0.099}} &{\bf{0.01897}} &{\bf{0.246}} &{\bf{0.716}} &1.906 & 1.329 & 0.0162 & 6.019 & {\bf{1.441}} &{\bf{4.686}} &4.915 & 4.633  \\
K2-MZvar-A09 & 75.6 & 0.109 & {\bf{0.01902}} &{\bf{0.247}} &{\bf{0.716}} &2.181 & 1.331 & 0.0171 & {\bf{6.006}} &1.431 & {\bf{4.757}} &{\bf{5.125}} &5.014  \\
\end{longtable}
\tablefoot{
         The numbers highlighted in bold indicate the values that satisfy the constraints listed in Table\,\ref{tab:targets}.
         The $\chitwo_{N=8}$ value is calculated using eight quantities ($\log\Lstar$, $\Teff$, and six from rms$(\delcs)$ to $\ABe$).
         The $\log\Lstar$ and $\Teff$ values are not listed here but well match observations in all the models.
         This table is available in electronic form on \href{https:doi.org/10.5281/zenodo.16789192}{Zenodo}.}

\end{appendix}

\end{document}